\documentclass[a4paper,12pt]{article}

\def\ad   {a^{\dagger}}

\def\al   {\alpha}

\def\HN {\hat N}

\def\HW {\hat W}

\def\HN {\hat N}
\def\HT {\hat T}

\def\CN {{\cal N}}
\def\HCN {\hat{\cal N}}

{\it}
{\it}
\def\expo{\rm {exp}}
\def\exp{\rm {e}}
{\it}
\def\Tr {\rm {Tr}}{\it}
\def\ogar {\overline\gamma_r}
\def\ogac {\overline\gamma_c}

\def\ga {\gamma}
\def\de {\delta}

\def\si {\sigma}

\def\la {\lambda}

\usepackage{times}
\usepackage{graphicx}
\begin{document}
\title{ About the sign ambiguity in the evaluation of Grand Canonical traces  
        for quasi-particle statistical density operators.} 
\author{G. Puddu\\
       Dipartimento di Fisica dell'Universita' di Milano,\\
       Via Celoria 16, I-20133 Milano, Italy}
\maketitle
\begin {abstract}
        A simple and general prescription for evaluating unambiguously
        the sign of the grand-canonical trace of quasi-particle statistical
        density operators (the so-called sign ambiguity in taking the
        square root of determinants) is given. Sign ambiguities of this kind
        appear in the evaluation of the grand-canonical partition function
        projected to good quantum numbers (angular momentum, parity and particle
        number) in the Hartree-Fock-Bogoliubov approximation at finite 
temperature, since traces
        are usually expressed as the square root of  determinants.
        A comparison is made with the numerical continuity method.
\par\noindent
{\bf{Pacs numbers}}: 21.60.Jz
\vfill
\eject
\end{abstract}
\section{ Introduction.}
         Nuclei at finite temperature are usually studied microscopically
         using the finite temperature Hartree-Fock (HF) or Hartree-Fock Bogoliubov (HFB)
         approximations (ref. [1],[2]), at least at first level of approximation.
         The basic quantity is the grand canonical partition function or
         equivalently the grand potential. Such a quantity contains an average
         of all possible contributions from different conserved quantum numbers such
         as angular momentum, parity and particle number. It is of course of great
         physical interest to study the partition function and related thermal averages
         using the partition function projected to the exact quantum numbers.
         This is especially true if the projection to good angular momentum is
         carried out without any assumption about axial symmetry, that is, if the full
         three dimensional angular momentum projector is used. As the temperature
         increases, we expect triaxial shapes to play a role and it is 
interesting 
         to see how shape transitions are obtained at different excitation energies.
\par
         Sometimes the full projector is replaced by a partial projector to good
         z-component of the angular momentum $J_z$. The partition function for a specified
         value of the angular momentum $J$ is then obtained by subtraction
         between the partition function at a $J_z$ and the partition function at
         $J_z+1$. This recipe, however, has a basic limitation that requires the exact
         evaluation of the partition function. This limitation is of course a problem
         when using even accurate approximations such as the HFB. Moreover, in the limit
         of $0$ temperature, the HFB ground state for a specified value of $J_z$ is
         not as accurate as the ground state obtained with a specified value of $J$.
         Also, the above recipe would pose severe problems for odd and odd-odd nuclei.
         Therefore, the use of the exact angular momentum projector is highly desirable.   
\par
         In the case of the the temperature dependent HFB approximation, a standard 
         result for the trace of the statistical
         density operator (the exponential of a quadratic form in the quasi-particle
         operators) states that it can be recast as a square root of a determinant (ref.[3]).
         This is a problem if the projected partition function is required, since
         an improper sign can lead to erroneous results. In the past the cure
         for this problem has been given with the so-called continuity argument.
         This argument states that the proper sign can be determined by constructing
         the statistical density operator from unity and then by determining the appropriate sign
         by imposing the continuity of the phase of the trace as we progressively rebuild
         the statistical density operator. This has been the recipe followed in ref. [3].
\par
         Recently this problem has been considered anew using the Grassmann algebra (ref.[4])
         for the determination of the sign for both overlaps of HFB wave functions and
         the trace of the statistical density operator (ref.[5]). Although the results obtained
         were not previously reported in the literature, in the case of the trace of the statistical density 
         operator, the sign ambiguity was not fully resolved since the vacuum contribution
         was still left as the square root of a determinant.
\par
         The purpose of this work is to show how all possible ambiguities can be
         resolved without referring to a numerical continuity argument, which may not be easy to implement.
         In the next section we shall derive the construction of the proper
         sign for the trace of the statistical density operator in rather general terms
         starting from  the properties  of the Lie algebra of the generators of the statistical density 
         operators as described in ref. [6]. Therefore the HFB approximation is only a special case of
         the recipe described below.
\bigskip
\section {Determination of the sign of the trace of the statistical density operator.}
\bigskip
\par\noindent{\it{$2.1 $ Symbols, definitions and basic properties.}}
\bigskip
\par
         As mentioned in the introduction we shall keep the discussion as general as
         possible. Let $N_s$ be the total number of the single particle states (that is
         neutrons plus protons). Let us consider an arbitrary antisymmetric complex 
         $2 N_s\times 2 N_s$ matrix 
         $A$ and let us define the row vector $\ga_r=(a,\ad)$, the
         collection of all annihilation and of all creation operators. In order to use
         consistently the matrix notations let us denote the column vector $\ga_c=col(a,\ad)$.
         A general statistical density operator (SDO for short) is written as
$$
\HW = \expo( {1\over 2} \ga_r A \ga_c)
\eqno(1)
$$
         No other limitations are imposed on this operator, except for the antisymmetry of the
         matrix $A$.
         Also let us define the $2N_s\times 2N_s$ matrix 
$$
\si = \left(\begin{array}{cc}
0 & 1\\
 1 & 0 \end{array}\right )
\eqno(2)
$$
         and the vectors
$$
\ogar = \ga_r \si \;\;\;\;\;\ \ogac=\si \ga_c
\eqno(3)
$$
         Eq. (1) can be recast as
$$
\HW = \expo( {1\over 2} \ogar S \ga_c),\;\;\;\;\;\;\;\ S=\si A
\eqno(4)
$$
        This is the form of the SDO  we shall use in the following.
        In the case of the HFB approximation  $S$ is hermitian.
        To any operator of the form of eq.(4) one can associate the matrix (without the caret)
$$
W = \exp^S 
\eqno(5)
$$
        Following ref. [6], the exponents of operators as in eq.(4) form a Lie algebra, and therefore 
        the product of any 
        two exponential operators of this form is an operator of the same form, moreover  the product 
        preserves   the association of eq.(5) that is, if
$$
\HW_1 \HW_2 =\HW
\eqno(6)
$$
        also
$$
W_1 W_2 = W
\eqno(7)
$$
        For example, $\HW_1$ is the rotation operator in terms of the Euler angles.
        In general, an arbitrary SDO is constructed as a product of
        several operators of the class of eq.(4), and for each one of them we know unambiguously
        the matrix $S$ of eq.(4). Eq. (6) tells us that the matrix $S$ corresponding to the product
        $\HW$ exists, but we cannot reconstruct this matrix from eq.(7) because of the $2\pi i$ ambiguity
        of the logarithm of the eigenvalues of $W$. This is the source of the sign ambiguity is the
        evaluation of the traces. From ref. [6], the operators of the class (4) transform the vectors
        $\ga$ in the following way
$$
\HW^{-1} \ga_c \HW = W \ga_c ,\;\;\;\;\;\; \HW^{-1} \ga_r \HW = \ga_r \tilde W, \;\;\;\;\;\;
\HW^{-1} \ogar \HW = \ogar W^{-1}
\eqno(8)
$$
        where $\tilde W$ denote the transpose of the matrix $W$.
        Moreover the following relation holds (cf. ref. [6])
$$
\si \tilde W \si = W^{-1}
\eqno(9)
$$
       It ensures that the transformed operators in eq. (8) obey the anticommutation relations.
\bigskip
\par\noindent\it{$2.2 $ The trace in the case the matrix $S$ is  known.}
\bigskip
\rm{}
\par
        If the matrix $S$ in eq.(5) is known, the grand canonical trace of $\HW$ can easily be evaluated
        without ambiguities. Let us prove it in the most general case, assuming $S$ is known. 
Consider a SDO
        $\HT$, then
$$
\Tr( \HW )=\Tr(\HT^{-1} \HW \HT)=\Tr ( \expo ({1\over 2} \HT^{-1} \ogar S \ga_c \HT)) \
=\Tr(\expo({1\over 2} \ogar T^{-1} S T \ga_c ))
\eqno(10)
$$
       where eq.(8) has been used. We shall prove that the matrix $T$ that diagonalizes $S$ 
       satisfies eq.(9).
\par\noindent
       It is easy to see that the eigenvalues of $S$ come in opposite pairs.
       In fact, the eigenvalue problem for $S=\si A$ written in the form ($T$ is the matrix of the 
       eigenvectors and the $\la$'s are the eigenvalues written in block form for convenience)
$$
S T = T \left(\begin{array}{cc}
 \la & 0\\ 0 & \la'\end{array}\right)
\eqno(11)
$$

       can be rewritten as 
$$
A \si (\si T \si) = (\si T \si) (\si \left (\begin{array}{cc}
 \la & 0\\
 0 & \la'\end{array}\right ) \si)
= (\si T \si) 
\left ( \begin{array}{cc} 
\la' & 0\\
 0 & \la \end{array}\right ) 
$$
       but $A \si= -\tilde S$ hence $ \la=-\la'$. The trace can now be trivially evaluated and the result is
$$
\Tr( \HW )= \exp^{ -{1\over 2} \sum_i^{N_s} \la_i} \prod_i^{N_s} ( 1+\exp^{\la_i})
\eqno(12)
$$
       provided one can show that $\HT$ is an element of the class of eq.(4), that 
is, 
       provided its associated matrix $T$ satisfies eq.(9)
       (which guarantees the legitimacy of the chain of steps in eq.(10)).
       In order to see this, consider the eigenvalue problem for $W$ written as
$$
W T = T 
\left( \begin{array}{cc}
 \exp^{\la} & 0\\
 0 & \exp^{-\la}\end{array}\right)
\eqno(13)
$$
       by applying at the left and at the right the matrix $\si$ ( $\si^2=1$) and 
       taking into account 
       eq.(9) one has
$$
( \tilde W )^{-1} \si T\si = \si T \si 
\left(\begin{array}{cc}
 \exp^{-\la} & 0\\
 0 & \exp^{\la}\end{array}\right)
\eqno(14)
$$
       Taking the inverse and the transpose of the above we obtain
$$
W  (\si \tilde T\si)^{-1} = (\si \tilde T\si)^{-1} 
\left(\begin{array}{cc}
 \exp^{\la} & 0\\
 0 & \exp^{-\la}\end{array}\right)
\eqno(15)
$$
       Therefore $T_{ik} $ and $(\si \tilde T\si)^{-1}_{ik} $ coincide apart a normalization constant $f_k\not=0$.
       If $f$ is the diagonal matrix of elements $f_k$ then
$$
f=(\si \tilde T\si)  T
\eqno(16)
$$
       Evaluating the above product one can show that $f$ has the doublet structure 
       $f=diag(f_1,f_2,...f_1,f_2,..)$.
       Therefore $ \sigma f^{-1/2} \sigma = f^{-1/2}$ and from this it follows that
       if $T$ does not satisfy eq. (9) the matrix $T/\sqrt{f}$ does.
\par
       Eq. (12) for the trace does not have any sign ambiguity since we had access to the eigenvalues of 
       $S$. If we do not have access to the matrix S, from eq. (12) taking the square, we have
$$
\Tr(\HW)^2 =  \det(1+W)
\eqno(17)
$$
       since we always have access to eigenvalues of $W$. This is where the sign ambiguity comes from.
       If we work with $W$ obtained from eq.(7), we never have access to the matrix $S$
       although we know it exists.
\bigskip
\par\noindent\it{ $ 2.3 $ The contribution of the vacuum.}
\bigskip
\rm{}
\par
       Consider now the fugacity dependent trace
$$
Z_{gc}(z)= \Tr ( \exp^{\al \HN} \HW )
\eqno(18)
$$
      where $z=\exp^{ \al}$ and $\HN$ is the particle number operator. For $z=0$ we isolate the contribution 
      of the vacuum. Let us define the
      operator of the class (4)
$$
\HCN=\expo {1\over 2}\ogar\left(\begin{array}{cc}
\al & 0 \\
0 & -\al \end{array}\right)\ga_c
\eqno(19)
$$
      and its associated matrix
$$
\CN= \left(\begin{array}{cc}
z & 0 \\
0 & 1/z \end{array}\right)
\eqno(20)
$$
      Let us set $ \HW(z)= \HCN\HW$. This operator has an associated matrix $W(z)=\CN W$. 
      Explicitly
$$
W=\left(\begin{array}{cc}
W_{11} & W_{12} \\
W_{21} & W_{22}\end{array}\right)
\eqno(21)
$$
    Then
$$
 Z_{gc}(z)^2=z^{N_s}\det(1+W(z))=\det\left(\begin{array}{cc}
1 & 0 \\
0 & z\end{array}\right)\det(1+\CN W)
\eqno(22)
$$
    This expression can be recast as
$$
 Z_{gc}(z)^2= \det (S_v+z S_p)=\det (S_v)\det (1+z S_v^{-1} S_p)
\eqno(23)
$$
   where 
$$
S_v=\left(\begin{array}{cc}
 1   & 0 \\
 W_{21} & W_{22}\end{array}\right),\;\;\;\;\;S_p=\left(\begin{array}{cc}
 W_{11}   & W_{12} \\
 0 & 1 \end{array}\right),\;\;\;\;\;
\eqno(24)
$$
     Since $Z_{gc}(z)$ is a polynomial in $z$ the eigenvalues of $S_v^{-1} S_p$ must come in degenerate 
     pairs
     $(\mu_k,\mu_k),\;\;k=1,,,N_s$. This argument is the similar to the one used in ref. [7].Therefore
$$
Z_{gc}(z)= \det(W_{22})^{1/2} \prod_{k=1}^{N_s}(1+z \mu_k)
\eqno(25)
$$
     It follows that, if we know the matrix $S$ (the log of $W$) then, using eq.(12),
$$
\det(W_{22})^{1/2} ={\exp^{ -{1\over 2} \sum_i^{N_s} \la_i} \prod_i^{N_s} ( 1+\exp^{\la_i})
\over \prod_k^{N_s}(1+\mu_k)}
\eqno(26)
$$
     Where we have set $z=1$. There are no sign ambiguities in $\prod_{k=1}^{N_s}(1+z \mu_k)$
     since its sign must be a continuous function of $z$. The only ambiguity is the 
$\det^{1/2}(W_{22})$
     and it is removed by eq.(12). Eq.(26) is the basic equation that allows us to remove the sign ambiguity
     also in the general case when we do not know the matrix $S$ and its eigenvalues unambiguously,
     as shown in the next subsection.
     Before leaving this section let us note that for $z=0$ we obtain the vacuum contribution
     to the Grand Canoncal trace of $\HW$
$$
<0| \HW | 0> = \det(W_{22})^{1/2}
\eqno(27)
$$
     where $|0>$ is the particle vacuum. 

\bigskip
\par\noindent{\it{$2.4 $ The trace in the general case.}}
\bigskip
\par
     Consider the SDO of the type
$$
\HW=\HW^{(b)}\HW^{(a)}
\eqno(28)
$$
    where $\HW^{(b,a)}$ are of the type of eq.(4), and let us assume we know explicitly
    the matrices $S^{(b)}$ and $S^{(a)}$. As previously mentioned, we do not know unambiguously the
    matrix $S$ associated with $\HW$, although we know the matrix $W$ associated with
    $\HW$, since $W=W^{(b)}W^{(a)}$. In what follows we shall need the following matrices
$$
D^{(b)}= W_{22}^{(b)-1} W_{21}^{(b)},\;\;\;\; C^{(a)}= W_{12}^{(a)} W_{22}^{(a)-1} 
\eqno(29)
$$
    Using eq.(25) (with $z=1$) we have
$$
Z_{gc}= (\det[W^{(b)} W^{(a)}]_{22})^{1/2} \prod_{k=1}^{N_s}(1+ \mu_k)
\eqno(30)
$$
  The square root of the determinant can be evaluated in the following way.
  Consider the  vacuum expectation value $<0| \HW^{(b)} \exp^{\al \HN} \HW^{(a)}|0>$. Then
  (cf. eq.(27))
$$
<0| \HW^{(b)} \exp^{\al \HN} \HW^{(a)}|0>= z^{N_s/2}(\det [ W^{(b)} \CN W^{(a)} ]_{22})^{1/2}
\eqno(31)
$$
    Direct  evaluation of the matrix product gives
$$
<0| \HW^{(b)} \exp^{\al \HN} \HW^{(a)}|0>=(\det [ W_{22}^{(b)} W_{22}^{(a)}+z^2 W_{21}^{(b)}W_{12}^{(a)} 
 ])^{1/2}
\eqno(32)
$$
    or 
$$
<0|\HW^{(b)} \exp^{\al \HN}\HW^{(a)}|0>=(\det W_{22}^{(b)})^{1/2} (\det 
W_{22}^{(a)})^{1/2}(\det(1+z^2 
D^{(b)} 
C^{(a)})^{1/2}
\eqno(33)
$$
    Since this vacuum contribution must be a polynomial in $z$, the eigenvalues of $D^{(b)}C^{(a)}$
    must come in degenerate pairs $(\nu_k,\nu_k)$.
    Therefore, considering only one eigenvalue for each degenerate pair,
$$
<0|\HW^{(b)} \exp^{\al \HN}\HW^{(a)}|0>=(\det W_{22}^{(b)})^{1/2} (\det W_{22}^{(a)})^{1/2}
\prod_{k=1}^{N_s}(1+z^2 \nu_k)
\eqno(34)
$$

    Finally setting $z=1$, we obtain for  the grand-canonical trace
$$
Z_{gc}=\det(W_{22}^{(b)})^{1/2} \det(W_{22}^{(a)})^{1/2}\prod_k(1+\nu_k)\prod_k(1+ \mu_k)
\eqno(35)
$$
     The only sign ambiguity in eq.(35) comes from the contributions of the two square roots.
     From eq.(34) (for $z=0$), one can see that each square root is again the vacuum contribution
     from $\HW^{(b)}$ and $\HW^{(a)}$,
     but we know already as to remove this ambiguity using eq. (26) for each $W^{(b)},W^{(a)}$,
     since we know the matrices $S^{(b)}$ and $S^{(a)}$.
     These considerations can easily be extended to a product of several SDO's.
\par
     As a final remark, it is possible (using eq. (9)) to prove that the matrices $ D^{(b)}$ and 
     $ C^{(a)}$ are antisymmetric and therefore these arguments amount to a quantum mechanical
     proof of the statement that the product of two antisymmetric matrices has eigenvalues
     in degenerate pairs (even $N_s$) or the odd one is zero (odd $N_s$). 
\bigskip
\par\noindent{\it{$2.5 $ A numerical test.}}
\bigskip
\par
     Essentially our method to fix the sign of the square root in the general case is based
     on an analytical continuity argument, supplemented by the fact that we know the
     contribution of the vacua of the factors $W^{(a)},W^{(b)},..$, because of eqs.(12) and (26).
     We performed extensive numerical tests of eq.(35), by considering an ensemble of $12\times 12$
     antisymmetric random matrices with matrix elements uniformly distributed in the interval
     $-a$ and $a$ ($a\simeq 3$). This random set generates the matrices $A^{(b)}$ and $A^{(a)}$
     and from eq.(4) the matrices $S^{(b)}$ and $S^{(a)}$. To test the above method we 
     consider the matrices $b S^{(b)}$ and  $b S^{(a)}$ with $b$ varying from $0$ to $1$ in 
     sufficiently  small steps, so that a numerical continuity argument can be 
     tested. 
     We also have used the conventional way of evaluating the grand canonical trace, by taking the square of
     eq.(35) and then numerically evaluating the square root and then following the phase of 
     this square root as $b$ is varied from $0$ to $1$. 
     More precisely, we evaluate the phase of the grand canonical trace and, if the phase changes between 
     consecutive values of $b$ by an amount larger than some value $\de \phi_{max}$ we change the phase by 
     $\pi$.
     In many cases the numerical continuity argument agrees with eq.(35) but in some case
     we found a sign disagreement. It is instructive to analyze these latter cases. In fig.(1) 
     we show the phase of eq.(35) as a function of $b$ for one of these instances, and in fig. (2) we show
     the phase for the same trace evaluated with the numerical continuity argument with two 
     different step sizes. For a relatively small number of steps ($N_{steps}=200)$,
     the phase shows a discontinuity, while for a much larger number of steps ($N_{steps}=800$),
     the  continuity of the phase as a function of $b$ is restored.
     The maximum phase change from one step to the next has been kept fixed for both cases
     to $\de \phi_{max}=0.5$.
     From fig.(2) we see that the source of the discontinuity is the rather large change in the phase
     for small variations of $b$, which is associated with a vertical slope of the phase.
     In this instance even $400$ steps fail to reproduce the continuity of the phase. This
     rather surprising result, was obtained because of the disagreement with eq.(35), otherwise
     it would have gone undetected. If we have to check the continuity of the phase for
     hundreds of cases (as it is in the case of the angular momentum projection), we can hardly
     check every single instance to insure the proper phase. If we consider smaller values 
     of $\de \phi_{max}$ we may need several thousands steps to restore continuity.
     This pathological behavior of the numerical continuity argument was found  in 
     presence of a vertical slope . 
\begin{figure}
  \begin{center}
    \includegraphics{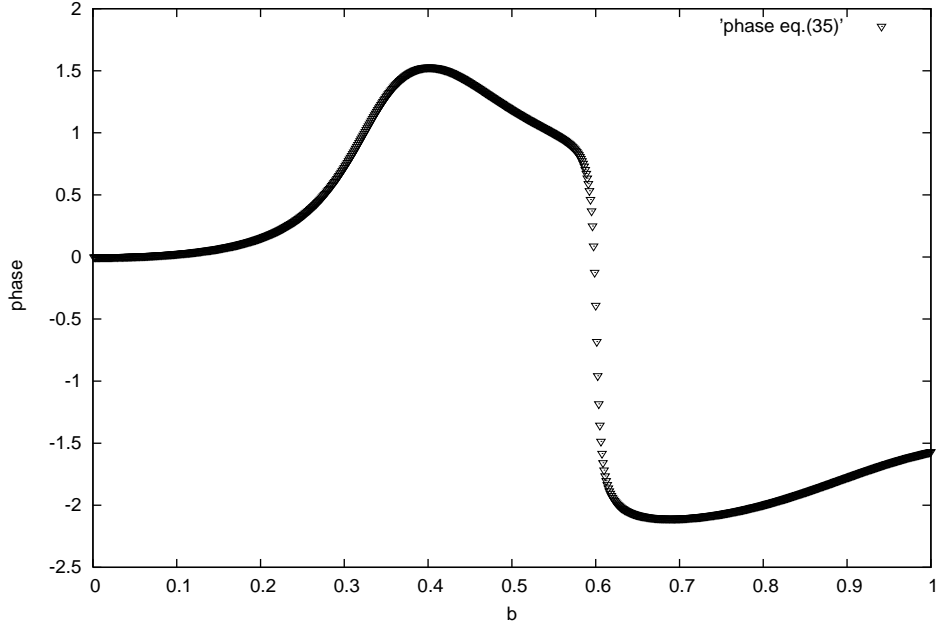}
  \end{center}
  \caption{The phase of eq.(35) as a function of $b$}
\end{figure}
\begin{figure}
  \begin{center}
    \includegraphics{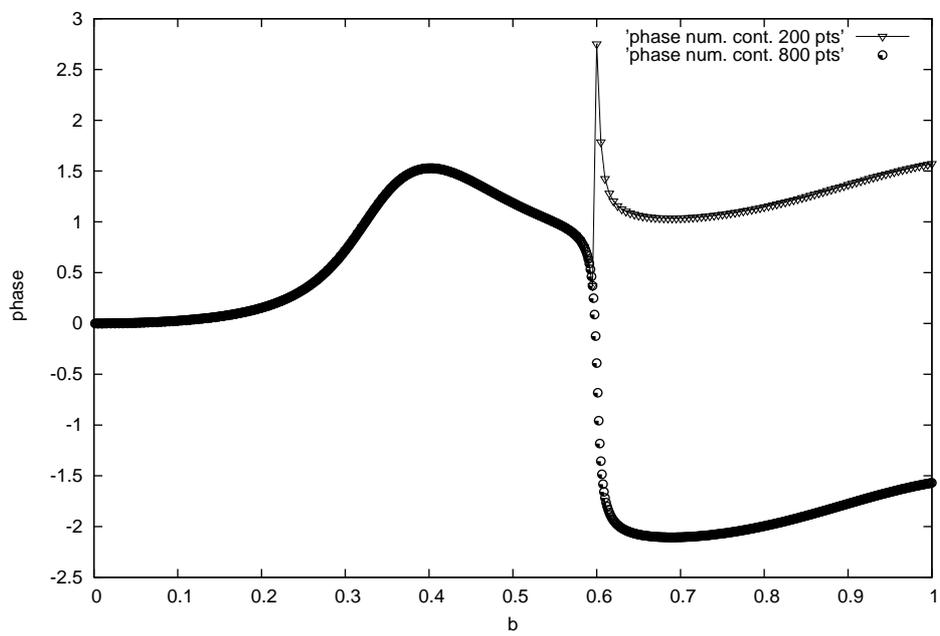}
  \end{center}
  \caption{The phase of the square root of the square of eq.(35) as a function of $b$, obtained with
  the numerical continuity method.}
\end{figure}
     This example does show the limitations of the numerical continuity method. 
\bigskip
\section {Conclusions.}
\bigskip
     In this work we have shown that the sign ambiguity in the evaluation of the trace
     of statistical density operator written as a product of elementary statistical density
     operators (this is usually the case in physical applications) can be effectively removed 
     by considering all factors separately,  evaluating the vacuum contributions using 
     eq.(12), then reconstructing the vacuum contribution of the full statistical density operator
     without taking any square root of determinant using eqs. (35).
     Quite surprisingly, a numerical continuity
     argument, computationally more involved, can fail to reproduce the proper sign in some 
     numerical test cases, unless we use a large number of steps.
     The recipe presented in this work opens the possibility to perform calculations of
     grand canonical partition functions within the HFB formalism with projectors to good
     quantum numbers.
\par

\vfill
\eject
\end{document}